\documentclass{article}
\usepackage[hyphens]{url} 
\usepackage{graphicx}
\usepackage[all]{xy}
\usepackage{hyperref}
\usepackage{amsmath,amsthm,amssymb}
\usepackage{cleveref}

\newtheorem{theorem}{Theorem}

\theoremstyle{definition}
\newtheorem{definition}[theorem]{Definition}
\newtheorem{lemma}[theorem]{Lemma}
\newtheorem{conjecture}[theorem]{Conjecture}
\newtheorem{remark}[theorem]{Remark}
\newtheorem{proposition}[theorem]{Proposition}

\usepackage{enumerate}
\usepackage{tikz}
\usetikzlibrary{automata,positioning}
\usetikzlibrary{decorations.pathmorphing}
\tikzset{snake it/.style={decorate, decoration=snake}}

\usepackage[utf8]{inputenc}
\usepackage{graphicx}
\usepackage{tikz}
\usepackage{enumerate}
\usetikzlibrary{matrix}
\usepackage{lscape}
\usetikzlibrary{automata,positioning}
\usepackage{caption}
\usepackage{subcaption}
\usepackage{multicol}
\usepackage{float}
\usepackage{makecell}
\usepackage{chngpage}
\usepackage{tikz-qtree}
\usepackage{mathtools}
\usepackage{tcolorbox}
\usepackage{collectbox}
\usepackage{pgf}
\usepackage{pgfplots}
\usepackage{longtable}
\usepackage[all]{xy}

\DeclareMathOperator{\VC}{VC}
\DeclareMathOperator{\TD}{T}

\DeclareMathOperator{\NFA}{NFA}
\DeclareMathOperator{\nfa}{nfa}
\DeclareMathOperator{\num}{int}
\DeclareMathOperator{\bin}{bin}

\newcommand{\abs}[1]{\lvert#1\rvert}
\renewcommand{\H}{\mathcal H}

\newcommand{\mc}{\makecell}

\newcommand{\mt}{\mathtt}
\begin{document}
\author{Bj{\o}rn Kjos-Hanssen\thanks{Corresponding author. Email: \textsf{bjoernkh@hawaii.edu}.
Address: Department of Mathematics, University of Hawai\textquoteleft i at M\=anoa, 2565 McCarthy Mall, Honolulu HI 96822, USA.},\\
Clyde James Felix, Sun Young Kim, Ethan Lamb, Davin Takahashi
}
\title{
	VC-dimensions of nondeterministic finite automata for words of equal length\thanks{
		The authors were supported by \emph{
			Faculty Mentoring Grants for Summer Undergraduate Research and Creative Works
		},
		sponsored by the
		Undergraduate Research Opportunities Program (UROP) in the
		Office of the Vice Chancellor for Research,
		University of Hawai\textquoteleft i.
		This work was partially supported by grants from the
		Simons Foundation (\#315188 and \#704836 to Bj\o rn Kjos-Hanssen).
	}
}
\maketitle
\begin{abstract}
Let $\NFA_b(q)$ denote the set of languages accepted by nondeterministic finite automata with $q$ states over an alphabet with $b$ letters.
Let $B_n$ denote the set of words of length $n$. We give a quadratic lower bound on the VC dimension of
\[
	\NFA_2(q)\cap B_n = \{L\cap B_n \mid L \in \NFA_2(q)\}
\]
as a function of $q$.

Next, the work of Gruber and Holzer (2007) gives an upper bound for the nondeterministic state complexity of finite languages contained in $B_n$,
which we strengthen using our methods.

Finally, we give some theoretical and experimental results on the dependence on $n$ of the VC dimension and testing dimension of $\NFA_2(q)\cap B_n$.

\end{abstract}
%\tableofcontents

	\section{Introduction}
		In this article we shall improve some results on the nondeterministic state complexity of finite languages, and investigate the VC dimension associated to fixed numbers of states for NFAs.
		Our methods build on work from the last three decades: by Ishigami and Tani \cite{MR1444188} for VC dimension of DFAs; Gruber and Holzer \cite{MR2362187} for nondeterministic state complexity; and Shallit and Wang \cite{MR1897300} and Hyde and Kjos-Hanssen \cite{MR3386523} for automatic complexity and languages consisting of words of the same length.
	\subsection{Dimension}

		The Vapnik--Chervonenkis dimension is an important tool in machine learning. We fix notation and give the definition:

		\begin{definition}
			Let $\H$ be a set family (a set of sets) and $C$ a set. Their \emph{intersection} is the set family
			$\H\cap C := \{H\cap C\mid H\in\H\}$.
			A set $C$ is \emph{shattered} by $\H$ if $\H\cap C$ contains all the subsets of $C$, i.e.:
			$|\H\cap C| = 2^{|C|}$.
			The \emph{VC dimension} of $\H$ is $\dim_{\VC}(\H)=\max\{\abs{C}: C\text{ is shattered by }\H\}$.
		\end{definition}
		The less-known \emph{testing dimension} was introduced by Kathleen Romanik at COLT'92 \cite{10.1145/130385.130422,MR1479623}
		and is the result of replacing an ``$\exists$'' by a ``$\forall$'':

		\begin{definition}
			Let $\mathcal Q$ be a concept class defined over a set $X$, i.e., a family of subsets of $X$.
			The \emph{testing dimension} of $\mathcal Q$ is
			\begin{align*}
				\dim_{\TD}(\mathcal Q, X)=\sup\{k\in\mathbb N \mid &\abs{X}\ge k\text{ and \emph{all}}\\ &\text{subsets of $X$ of cardinality $k$ are shattered by $\mathcal Q$}\}.
			\end{align*}
		\end{definition}
		Unlike VC dimension, testing dimension on its face depends on $X$ as well as $\mathcal Q$.
		We typically have $X=\bigcup\mathcal Q$, but note \Cref{thm:depends} below.

\begin{figure}
			\includegraphics[width=12cm]{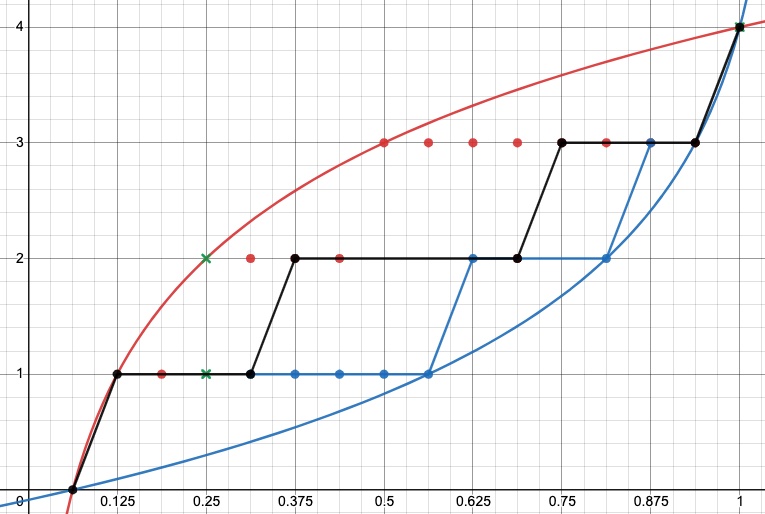}
			\begin{itemize}
				\item Upper bound $4+\log_2 x$ for $\dim_{\VC}$ in red.
				\item Lower bound $-\log_2(1+2^{-4}-x)$ for $\dim_{\TD}$ in blue from \Cref{rem:TD-lb}.
				\item Lower bound $x=2^{-4}+\sum_{k=0}^{y-1}\binom{4}k 2^{-4}$ for $\dim_{\VC}$ from \Cref{Sauer} in black.
				\item $\dim_{\VC}, \dim_{\TD}$ of $\NFA'_2(q)\cap B_2$ for $q=1,2$ in green.
			\end{itemize}
			\caption{VC and testing dimension of $\NFA'_b(q)\cap B_2$: upper and lower bounds.
			$y=\dim_{\VC}(\mathcal C)$ as a function of $x=c/16$,
			where $\abs{\mathcal C}=c\le 2^{2^2}$ is a set family over $B_2$.
			}\label{desmos-VC-2}
		\end{figure}

\begin{figure}
			\includegraphics[width=12cm]{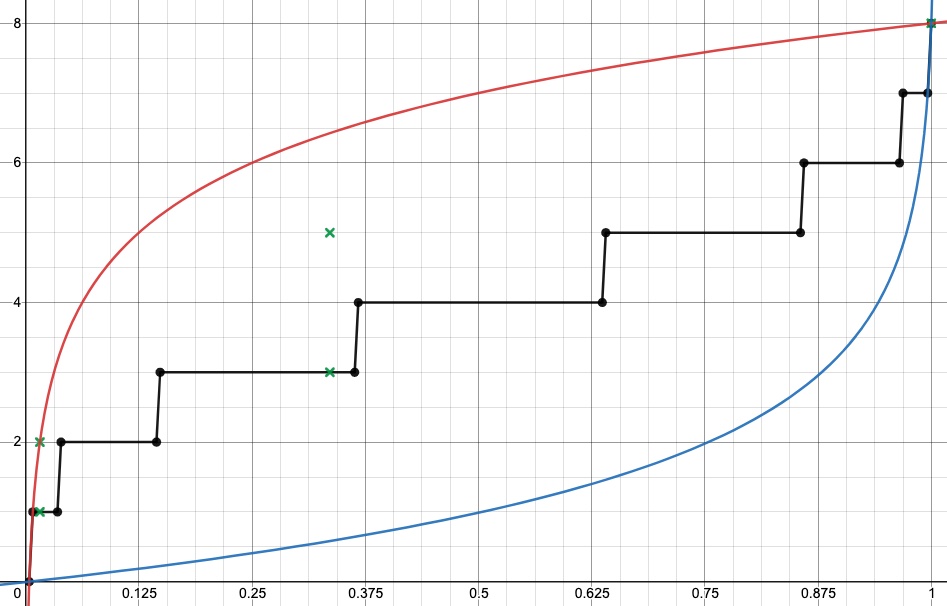}
			\begin{itemize}
				\item Upper bound $2^{3}+\log_2 x$ for $\dim_{\VC}$ in red.
				\item Lower bound $-\log_2(1+2^{-{2^3}}-x)$ for $\dim_{\TD}$ in blue.
				\item Lower bound $x=2^{-{2^3}}+\sum_{k=0}^{y-1}\binom{4}k 2^{-{2^3}}$ for $\dim_{\VC}$ from \Cref{Sauer} in black.
				\item $\dim_{\VC}, \dim_{\TD}$ of $\NFA'_2(q)\cap B_3$ for $q=1,2,3$ in green.
			\end{itemize}
			\caption{VC and testing dimension of $\NFA'_b(q)\cap B_3$: upper and lower bounds.
			$y=\dim_{\VC}(\mathcal C)$ as a function of $x=c/2^{2^3}$, where $\abs{\mathcal C}=c\le 2^{2^3}$ and $\mathcal C$ is a set family over $B_3$.
			}\label{desmos-VC-3}
		\end{figure}

		A visualization of the relationship between $\dim_{\TD}$ and $\dim_{\VC}$ over $B_n$ for $n=2$ and $n=3$ can be found in \Cref{desmos-VC-2,desmos-VC-3}.
		There, we make use of \Cref{Sauer,thm:TD-lb}.
		\begin{theorem}[Sauer--Shelah Lemma \cite{MR307902,MR307903}]\label{Sauer}
			There is no family of subsets of $U$, $u=\abs{U}$, of cardinality $>\sum_{k=0}^m\binom{u}k$ with VC dimension $\le m$.
		\end{theorem}

		\begin{theorem}\label{thm:TD-lb}
			Let $c$ and $u$ be nonnegative integers, and let $U$ be a set of cardinality $u$. The following are equivalent:
			\begin{enumerate}
				\item For each set family $\H$ over $U$, $\abs{\H}>c\implies\dim_{\TD}(\H)\ge m$.
				\item $c\ge 2^{u}-2^{u-m}$.
			\end{enumerate}
		\end{theorem}
		\begin{proof}
			(2) $\implies$ (1):
			Assume (2), let $\H$ be given, and let $D\subseteq C\subseteq U$ with $\abs{C}=m$. We must show that there is some
			$H\in\H$ such that $H\cap C=D$. The number of sets $H$ with $H\cap C=D$ is $2^{u-m}$.
			By (2), $\abs{\H}>2^u-2^{u-m}$, hence the complement $\H^c$ of $\H$ satisfies $\abs{\H^c}=2^u-\abs{\H}< 2^{u-m}$.
			Hence $\{H:H\cap C=D\}\not\subseteq\H^c$, as desired.

			(1) $\implies$ (2):
			Suppose $c<2^{u}-2^{u-m}$. Let $D\subseteq C\subseteq U$ with $\abs{C}=m$. Let $\H=\{H: H\cap C\ne D\}$.
			Then $\abs{\H}=2^u-2^{u-m}>c$, and $C$ and $D$ witness that $\TD(\H)<m$.
		\end{proof}
		\begin{remark}\label{rem:TD-lb}
			By \Cref{thm:TD-lb}, letting $x=(c+1)2^{-u}\ge 1-2^{-m}+2^{-u}=1-2^{-\TD}+2^{-u}$ we obtain a lower bound used in \Cref{desmos-VC-2,desmos-VC-3}.
			A basic and obvious upper bound on $\dim_{\VC}$ used there is: $\dim_{\VC}(\mathcal H)\le \log_2\abs{\mathcal H}$.
		\end{remark}
	\subsection{Automata}

		For a nonnegative integer $k$, we let $[k]=\{0,\dots,k-1\}$. Thus $[2]^*=\{0,1\}^*$ is the set of all finite binary words.

		\begin{definition}
			Let $\nfa'_b(q)$ ($\nfa_b(q)$) denote the class of all nondeterministic finite automata with $q$ states and 1 accept state (arbitrary number of accept states) over an alphabet of cardinality $b$. 
			Let $\Sigma$ be a set. Then $\nfa_{\Sigma}(q)$ is the set of all NFAs with $q$ states over the alphabet $\Sigma$.

			The language accepted by the automaton $M$ is $L(M)$.
			Let
			\[
				\NFA_b(q)=\{L(M)\mid M\in \nfa_b(q)\},
			\]
			$\NFA_\Sigma(q)=\{L(M)\mid M\in \nfa_{\Sigma}(q)\}$, and similarly define $\NFA'_b(q)$ and $\NFA'_{\Sigma}(q)$.
		\end{definition}

		For example, $\NFA_b(q)\cap B_n$ is a set of ``slices'' of languages accepted by $q$-state NFAs, where $B_n=\{0,1\}^n$.

		\begin{theorem}\label{thm:depends}
			Let $\mathcal S=\bigcup_{q=1}^\infty\NFA_{\{\mt 0\}}(q)$. Then
			\[
				\dim_{\VC}(\mathcal S)=\infty, \text{ but }
				\dim_{\TD}(\mathcal S,\{\mt 0,\mt 1\}^*)=0.
			\]
		\end{theorem}
		\begin{proof}
			$\mathcal S$ shatters $\{\mt{0}^k:k\le n\}$ for any $n$,
			but does not shatter $\{\mt{1}\}$.
		\end{proof}
	\section{Main results}

		Gruber and Holzer \cite{DBLP:conf/dcfs/GruberH06,MR2362187} gave the following construction. Intuitively, we have states $p_w$ indicating that we have seen the symbols in $w$, and
		states $q_w$ indicating that the symbols in $w$ remain to be seen.
		
		\begin{definition}\label{gh-def}
			Let $\lambda$ denote the empty word.
			Let $L\subseteq B_n$ or $L\subseteq \{0,1\}^{\le n}$.
			Let $\ell = \lfloor (n-1)/2\rfloor$ and $m=\lceil (n-1)/2\rceil$.
			We construct a nondeterministic finite automaton $A=(Q,\{0,1\},\delta,p_{\lambda},F)$, where $Q=P_1\cup P_2$ (disjoint union)
			with $P_1=\{p_w \mid w\in\{0,1\}^* \text{ and }\abs{w}\le\ell\}$
			and  $P_2=\{q_w \mid w\in\{0,1\}^* \text{ and }\abs{w}\le   m\}$,
			and the set of final states $F=\{q_{\lambda}\}\cup \{p_{\lambda} \mid \lambda \in L\}$. (Note that $1\le \abs{F}\le 2$, and $\abs{F}=2\iff \lambda\in L$.)
			The transition function is specified as follows:
			\begin{enumerate}
				\item For all $p_w\in P_1$ and $a\in\{0,1\}$, if $\abs{w}<\ell$ then the set $\delta(p_w,a)$ contains the element $p_{wa}$.
				\item For all $w\in L\setminus\{\lambda\}$, if $w=xay$ is the unique decomposition where $\abs{x}=\lfloor (\abs{w}-1)/2\rfloor$,
				$a$ is a single letter, and $\abs{y}=\lceil (\abs{w}-1)/2\rceil$, then let $\delta(p_x,a)$ contain the element $q_y$.
				\item For all $q_w\in P_2\setminus\{q_{\lambda}\}$ and $a\in\{0,1\}$, the set $\delta(q_{aw},a)$\footnote{
				Gruber and Holzer's papers have instead the typo $\delta(p_{aw},a)=q_w$, which we correct here.
				} contains the element $q_w$.
			\end{enumerate}
		\end{definition}

		As a first approximation to the construction we will employ in \Cref{thm:HGH}, let us restate \Cref{gh-def} for the case $L\subseteq B_n$ with $n$ an odd number.
		\begin{definition}\label{def:gh-odd}
			Let $L\subseteq B_n$, where $n=2m+1$ is odd.
			We construct a nondeterministic finite automaton $A=(Q,\{0,1\},\delta,p_{\lambda},F)$, where $Q=P_1\cup P_2$ (disjoint union)
			with $P_1=\{p_w \mid w\in\{0,1\}^* \text{ and }\abs{w}\le m\}$
			and  $P_2=\{q_w \mid w\in\{0,1\}^* \text{ and }\abs{w}\le m\}$,
			and $F=\{q_{\lambda}\}$.
			The transition function is specified as follows:
			\begin{enumerate}
				\item For all $p_w\in P_1$ and $a\in\{0,1\}$, if $\abs{w} < m$ then the set $\delta(p_w,a)$ contains the element $p_{wa}$.
				\item For all $w\in L$, if $w=xay$ is the unique decomposition where $\abs{x}=\abs{y}=m$ and
				$a$ is a single letter, then let $\delta(p_x,a)$ contain the element $q_y$.
				\item For all $q_w\in P_2\setminus\{q_{\lambda}\}$ and $a\in\{0,1\}$, the set $\delta(q_{aw},a)$ contains the element $q_w$.
			\end{enumerate}
		\end{definition}

		If $x$ and $y$ are words and $x$ is a prefix of $y$, so that $y=xz$ for some word $z$, then we write $x\preceq y$.
		We denote the reversal of $x$ by $x^R$. If $x$ is a suffix of $y$ then consequently we may write $x^R\preceq y^R$.

		Our strategy for obtaining lower bounds for VC dimension in \Cref{thm:HGH} will be to remove some states $p_w$, and then to also remove any state $p_v$ where all $p_w$ with $w\succeq v$ are removed. The counting of the states removed this way will turn out to hinge on \Cref{def:a}.
		
		\begin{definition}\label{def:a}
			The function $a$ is defined by $a(0)=0$ and for $n\ge 0$,
			$a(n+1)=a(n)+1+t$ where $n+1=(2k+1)2^t$ for some $k$.
		\end{definition}
		The first few values of $a$ are tabulated below and at \cite{A005187}.

		\begin{tabular}{r|r r r r r r r r}
		$r$ &1&2&3&4&5& 6& 7& 8\\
		$a(r)$&1&3&4&7&8&10&11& 15
		\end{tabular}

		\Cref{lem:arndt} was stated, but not proved, by J\"org Arndt in \cite{A005187}.
		\begin{lemma}\label{lem:arndt}
			$a(n) = 2n - w(2n)$ where $w(n)$ is the binary Hamming weight of $n$.
		\end{lemma}
		\begin{proof}
			By induction.
			\emph{Base case:} For $n=0$, $a(0)=0=0-0=2\cdot-w(2\cdot 0)$, as desired.

			\emph{Induction step:} Assume $a(n)=2n-w(2n)$.
			By definition, $a(n+1)=a(n)+1+t$ where $n+1=(2k+1)2^t$ for some $k$.
			The Hamming weight of
			\[
			2(n+1) = (2k+1)2^{t+1}
			\]
			compared to that of
			\[
			2n = 2((2k+1)2^t-1) = (2k+1)2^{t+1}-2
			\]
			is $t-1$ smaller, i.e.,
			\begin{equation}\label{eq:above-analysis}
				w(2(n+1))+t-1=w(2n).
			\end{equation}
			So then
			\begin{align*}
				a(n+1) 	&= a(n)					&+ 1 + t	&\quad\text{(by definition)}\\
						&= 2n - w(2n)           &+ 1 + t &\quad\text{(by the induction hypothesis)}\\
						&= 2n - (w(2(n+1))+t-1) &+ 1 + t &\quad\text{(by \eqref{eq:above-analysis})}\\
						&= 2(n+1) - w(2(n+1)).	& \qedhere
			\end{align*}
		\end{proof}

		\begin{lemma}\label{lem:arndt2}
			$a(n)\ge 2n - \lceil\log_2 (n+1)\rceil$ for all $n$.
		\end{lemma}
		\begin{proof}
			Let $w(n)$ be the binary Hamming weight of $n$.
			Note that $w(n)\le w(2^s-1)=s$ for each integer $s\ge 0$ and $n<2^s$.
			In particular, letting $s=\lceil\log_2(n+1)\rceil$,
			we have $n < n+1 \le 2^s$, and hence
			$w(n)\le w(2^{\lceil\log_2 (n+1)\rceil}-1) =  \lceil\log_2 (n+1)\rceil$ for each $n\ge 0$.
			By \Cref{lem:arndt} we are done.
		\end{proof}

		We denote the binary numerical value $\sum w_i 2^{m-i}$ of a word $w=w_1\dots w_m$, where each $w_i\in\{0,1\}$, by $\num_2(w)$. For example,
		\[
			\num_2(00)=0, \num_2(01)=1, \num_2(10)=2, \num_2(11)=3.
		\]
		For any integer $0\le k<2^n$, we write $\bin_n(k)$ for the word $w$
		of length $n$ with $\num_2 w=k$. For example, $\bin_4(3)=0011$.

		%For a set of words $E$, we write $E^{\preceq,n}=\{w\in B_n: \exists e\in E, e\preceq w\}$, and $E^R=\{w^R\mid w\in E\}$.
		\begin{lemma}\label{lem:a-prime}
			Let $E_r=\{w\in B_m: \num_2(w)<r\}$,
			\[
				P_r = \{w\in [2]^{\le m}\mid \forall v\in B_m, w\preceq v\implies v\in E_r\},
			\]
			and
			\(
				a'(r)=\abs{P_r}.
			\)
			Then $a'=a$ from \Cref{def:a}.
		\end{lemma}
		\begin{proof}
			Since $E_0=\emptyset$ and each $w\in [2]^{\le m}$ has some $v\in B_m$ with $w\preceq v$, we have
			\[
				a'(0)=\abs{\{w\in [2]^{\le m}\mid \forall v\in B_m, w\preceq v\implies v\in E_0\}}=0.
			\]
			For each removed word in $E_r$ we remove it from $P_r$,
			and additionally, for the $(2k+1)2^t$th removed word from $E_r$
			we remove $t$ more words from $P_r$. For instance, when the 4th word is removed from $E_r$ we remove 3 words from $P_r$.
			Thus, $a'(n+1)=a'(n)+1+t$ where $n+1=(2k+1)2^t$ for some $k$ as desired.
		\end{proof}

		For functions $f(q), g(q)$ we write $f\lesssim g$ if for all $\epsilon>0$,
		$f(q)\le (1+\epsilon)g(q)$ for all large enough $q$.

		\Cref{thm:HGH} is our main theorem, and uses a modification inspired by Hyde and Kjos-Hanssen \cite{MR3386523} of the Gruber--Holzer construction to get information about VC dimension over $B_n$.
		\begin{theorem}\label{thm:HGH}
			Let $n\ge 0$ and $q\ge 1$ be integers. Then
			\[
				\dim_{\VC}(\NFA'_2(q)\cap B_n)\gtrsim q^2/4.
			\]
		\end{theorem}
		\begin{proof}
			Let $m=\lfloor n/2\rfloor$, so that $n\in\{2m,2m+1\}$.
			Let $r$ be the least integer such that $2^{m+1}-1-a(r)\le q$.
			Let $\tilde q=2^{m+1}-1-a(r)\le q$.
			Let $W=\{w\in B_m\mid \num_2(w)\ge 2^m-r\}$.
			We claim that we can shatter the following sets $S$ using $\NFA'_2(q)$: 
			\begin{align*}
				 n=2m+1&\implies S=\{wb'w^R \mid w\in W, b'\in [2]\}.\\
				n=2m, r\le 2^{m-1}&\implies S=\{wb'v \mid w\in W, b'\in [2], \abs{v}=m-1\}.\\
				n=2m, r>2^{m-1}   &\implies S=\{wb'v \mid w\in W, b'\in [2], \abs{v}=m-1, 1v^R\in W\}.
			\end{align*}
			Thus, let $S'\subseteq S$. We define an automaton $M'$ that accepts every words in $S'$ and rejects every words in $S\setminus S'$.
			To do so, it suffices to
			\begin{itemize}
				\item construct an automaton $M$ with $\tilde q$ states and $L(M)\cap B_n=S'$ (hence $L(M)\cap S=S'$) and then
				\item extend $M$ to an automaton $M'$ with $q$ states and $L(M)=L(M')$, by adding $q-\tilde q$ inaccessible states.
			\end{itemize}
			Thus we may assume for notational convenience that $\tilde q=q$.

			\emph{Construction of $M$}.
			The set of states is
			\[
				Q=\{p_v\mid \exists w\succeq v, \abs{w}=m, \num_2(w)\ge r\},
			\]
			and contains some states $q_w$ that are identified with (equal to) states of the form $p_v$ as follows.
			\begin{itemize}
			\item If $n$ is odd then $q_w=p_{w^R}$ for each $w$ with $\abs{w}\le m$ and $p_{w^R}\in Q$.

			\item If $n$ is even, let $q_w=p_{1w^R}$ for each $w$ with $\abs{w}\le m-1$.
			\end{itemize}

			The set of final states is
			\[
				F=\begin{cases}
					\{p_\lambda\}&\text{if $n$ is odd},\\
					\{p_1\}&\text{if $n$ is even}.
				\end{cases}
			\]
			%(The choice of $p_1$ instead of $p_0$ here was arbitrary.)
			The transition function $\delta$ is defined by:
			
			1. For $w$ with $\abs{w}=m$ and $p_w\in Q$, $\delta(p_w,b')=\{q_v\mid  wb'v\in S'\}$.
			
			2. For other inputs, $\delta$ is as in \Cref{def:gh-odd}.

			This completes the construction of $M$. For examples see \Cref{fig:HGH-even,fig:HGH-odd}.

			By a parity consideration only the intended paths of length $n$ lead to the accept state, hence $L(M)\cap B_n=S'$.

			Let $E=\{w\in B_m: \num_2(w)<r\}$ and $N=B_m\setminus E$.
			Note that $a'(r)=\abs{\{w\mid \forall v\in B_m, w\preceq v\implies v\in E\}}$ is the number of words such that all its extensions to length $m$ are in $E$. So
			\begin{align*}
				q &= 2^{m+1}-1 - a'(r) &\\
				  &= 2^{m+1}-1 - a(r) & \text{by \Cref{lem:a-prime}}\\
				 &< 2^{m+1} - a(r) &\\
				&\le 2^{m+1}-2r + \lceil\log_2 (r+1)\rceil & \text{by \Cref{lem:arndt2}} \\
				&\sim 2(2^m - r). &
			\end{align*}

			\emph{Case 1: $n$ is odd.}
			The lower bound on VC dimension witnessed by $M$ is then
			\begin{eqnarray*}
				\dim_{\VC}(\NFA'_2(q)\cap B_n)
				&\ge&\abs{N}\times\abs{\{0,1\}}\times\abs{N}\nonumber \\
				&=&(2^m-r)\times 2\times (2^m-r)=2(2^m-r)^2\\
				&\gtrsim& 2(q/2)^2 = q^2/2.
			\end{eqnarray*}
			Asymptotically the VC dimension achievable $d$ for $n=2m+1$ length satisfies
			\begin{equation*}\label{eq:fits}
				d\gtrsim 2(q/2)^2 = q^2/2.
			\end{equation*}

			\emph{Case 2: $n$ is even.}
			There are two subcases, $r\le 2^{m-1}$ and $r>2^{m-1}$.

			Subcase $r\le 2^{m-1}$:
			Instead of $d=2(2^m-r)^2$ we get $d=2(2^{m-1})(2^m-r)$, and $q\lesssim 2(2^m-r)$.
			In particular $d\ge q 2^{m-1} \ge q^2/4$.

			Subcase $r>2^{m-1}$:
			Then $d=2(2^m-r)^2$
			which is the same expression in terms of $m$ as for the odd case.
		\end{proof}

		\Cref{thm:HGH} can be compared with a result from the literature:
		\begin{theorem}[{Ishigami and Tani ALT'93 \cite[Theorem 4.1]{MR1293747}}]\label{thm:fitting} %was not included in their later journal paper:
			Let $\Sigma$ be a finite alphabet, $b=\abs{\Sigma}$, and let $q\ge 1$ be an integer. Then
			\[
				(b-1)q^2\le \dim_{\VC}(\NFA_b(q))\le bq^2.
			\]
		\end{theorem}

		\begin{remark}
			The construction in \Cref{thm:HGH} is sharp for the number of states in the case $q=3$, $n=3$.
			Consider the set $\{000,011,110,111\}$.
			There is a 3-state solution produced by our construction:

			\begin{tikzpicture}[shorten >=1pt,node distance=2.5cm,on grid,auto]
				\node[state, initial above, accepting] (s) {};
				\node[state] (s0) [below left =of s] {};
				\node[state] (s1) [below right=of s] {};
				\path[->]
					(s0) edge [loop left] node {$0$} (s0)
					(s1) edge [loop right] node {$1$} (s1);
				\path[<->]
					(s) edge node [above] {$0$} (s0)
					(s) edge node {$1$} (s1)
					(s0) edge node [below] {$1$} (s1);
			\end{tikzpicture}
		\end{remark}

		The nondeterministic state complexity of a language $L$ is denoted $\mathrm{nsc}(L)$.
		\begin{remark}\label{rem:gh-wrong}
			Gruber and Holzer {\cite[Lemma 12]{MR2362187}} claimed that for all $L\subseteq\{0,1\}^{\le n}$,
			$\mathrm{nsc}(L)<\frac3{\sqrt 2}\sqrt{2^n}$. The proof in \cite{MR2362187}
			contains the following mistake.
			If $\ell=\lfloor (n-1)/2\rfloor$ and $m=\lceil (n-1)/2\rceil$ then they claimed that
			\begin{equation*}\label{eq:gh-wrong}
				2^{\ell+1}-1+2^{m+1}-1<\frac3{\sqrt 2}\sqrt{2^n}.
			\end{equation*}
			A correct estimate would be
			\begin{align*}\label{eq:gh-corrected-even}
				2^{\ell+1}-1+2^{m+1}-1
				&= 
				\begin{cases}
				3\sqrt{2^n}-2 & \text{if $n=2m$},\\
				2\sqrt 2\cdot \sqrt{2^{n}}-2&\text{if $n=2m+1$},
				\end{cases}
				\\
				&\le 3\sqrt{2^n} - 2.
			\end{align*}
		\end{remark}
		We state \Cref{lem:gh} in a corrected form.
		\begin{theorem}[Gruber and Holzer {\cite[Lemma 12]{MR2362187}}]\label{lem:gh}
			Assume $L\subseteq\{0,1\}^{\le n}$.
			Then $\frac12\sqrt{2^n}<\mathrm{nsc}(L)<3\sqrt{2^n}$.
		\end{theorem}
		\begin{theorem}
			Assume $L\subseteq B_n$.
			Then $\mathrm{nsc}(L)< 2\sqrt{2^n}$.
		\end{theorem}
		\begin{proof}
			Let $m$ be such that $n=2m+1$ or $n=2m$, i.e., $m=\lfloor n/2\rfloor$.
			The number of states required in \Cref{thm:HGH} with $r=0$ is
			\[
				 2^{m+1}-1 \le 2^{n/2+1}-1.\qedhere
			\]
		\end{proof}

	\section{Increasing the word length}

		\Cref{tab:vc-td} contains some results of exhaustive search by computer as well as some consequences of our theoretical results. Looking at the (incomplete) table suggests the following.
		\begin{conjecture}\label{cocoaRef}
			For all integers $n\ge 0$ and $q\ge 1$,
			\[
				\dim_{\VC}({\NFA'_2(q)}\cap B_n) \le \dim_{\VC}({\NFA'_2(q)}\cap B_{n+1}).
			\]
 		\end{conjecture}
		Informally, \Cref{cocoaRef} says that
		automata with a fixed number of states can shatter sets of \emph{long} words just as well as they can shatter sets of \emph{short} words.
		In particular, we do not know whether
		\[
			\dim_{\VC}(\NFA'_2(3)\cap B_4) \le^? \dim_{\VC}(\NFA'_2(3)\cap B_5).
		\]
		Resolving this would take about a month of continuous computation with our current code and computer. Although stated as a conjecture, we suspect the negation of \Cref{cocoaRef} is true and that for $n=6$ and $q=7$, $q$-state NFAs cannot be used to shatter a set of size $32$. In the remainder of the paper we prove some results related to \Cref{cocoaRef}. %June 27, 2021.

		\begin{theorem}\label{thm:may 26 2021}
		For all integers $n\ge 0$ and $q\ge 1$,
			\[
				\min(2^n,\dim_{\TD}({\NFA_2(q)}\cap B_{n+1})) \le \dim_{\TD}({\NFA_2(q)}\cap B_n).
			\]
		\end{theorem}
		\begin{proof}
			Let $s=\min(2^n,\dim_{\TD}({\NFA_2(q)}\cap B_{n+1}))$.
			Since $s\le 2^n$, to show
			\[
				\dim_{\TD}({\NFA_2(q)}\cap B_n)\ge s,
			\]
			given $S\subseteq B_n$ with $\abs{S}=s$,
			we must demonstrate how to shatter $S$ using $\NFA_2(q)$.
			Let $S'\subseteq S$.
			Let $T=S0=\{\sigma 0:\sigma\in S\}$ and similarly $T'=S'0$.
			Since $T\subseteq B_{n+1}$ and $\abs{T}=s\le \dim_{\TD}({\NFA_2(q)}\cap B_{n+1})$,
			there exists $M\in\nfa_2(q)$ with $T'\subseteq L(M)$, $(T\setminus T')\cap L(M)=\emptyset$.
			Let $N\in\nfa_2(q)$ be obtained from $M$ by letting the final states of $N$ be exactly those of $M$ that have a transition using 0 to a final state of $M$.
			Then $S'\subseteq L(N)$ and $(S\setminus S')\cap L(N)=\emptyset$, as desired.
		\end{proof}
		We do not know whether \Cref{thm:may 26 2021} is true with $\NFA'$ in place of $\NFA$; see \Cref{tab:vc-td}.
		\begin{theorem}
			For all integers $n\ge 0$ and $q\ge 1$,
			\begin{eqnarray*}
				\dim_{\VC}({\NFA_2(q)}\cap B_{n+1})=2^{n+1} &\implies& \dim_{\VC}({\NFA_2(q)}\cap B_n)=2^n, \quad\text{and}\\
				\dim_{\TD}({\NFA_2(q)}\cap B_{n+1})=2^{n+1} &\implies& \dim_{\TD}({\NFA_2(q)}\cap B_n)=2^n.
			\end{eqnarray*}
		\end{theorem}
		\begin{proof}
		Since $\dim_{\VC}({\NFA_2(q)}\cap B_{n+1})=2^{n+1}$, we also have $\dim_{\TD}({\NFA_2(q)}\cap B_{n+1})=2^{n+1}\ge 2^n$.
		By \Cref{thm:may 26 2021}, $\dim_{\TD}({\NFA_2(q)}\cap B_n)\ge 2^n$.
		Since $\dim_{\TD}\le\dim_{\VC}$ in general, this gives $\dim_{\VC}({\NFA_2(q)}\cap B_n)\ge \dim_{\TD}({\NFA_2(q)}\cap B_n)\ge 2^n$ as desired.
		\end{proof}

		\begin{theorem}
			For all integers $n\ge 0$ and $q\ge 1$,
			\[
				\dim_{\VC}({\NFA'_2(q)}\cap B_n)\le \dim_{\VC}(\NFA'_2(q+1)\cap B_{n+1}).
			\]
		\end{theorem}
		\begin{proof}
			Let $S\subseteq B_n$ be a set that is shattered by $\NFA'_2(q)$.
			Let $T=\{0\sigma \mid \sigma\in S\}$. Then $\abs{T}=\abs{S}$ and $T$ is shattered by $\NFA'_2(q+1)$.
		\end{proof}

		\begin{definition}
			For a fixed $q$ let us say that two sequences of words $x_1,\dots,x_k$ and $y_1,\dots,y_k$ are $\approx_q$-similar if
			for each $F\subseteq [k]$ there is an automaton $M$ with $q$ states such that
			\begin{eqnarray*}
				\{x_i\mid i\in F\}\cup \{y_i\mid i\in F\}&\subseteq& L(M),\quad\text{ and}\\
				\left(\{x_i\mid i\not\in F\}\cup \{y_i\mid i\not\in F\}\right) &\cap& L(M)=\emptyset.
			\end{eqnarray*}
			Let us say that two sets of words are $\sim_q$-similar if there exist orderings of them that are $\approx_q$-similar.
		\end{definition}
		For instance, $\{\mt{0}^n,\mt{1}^n\}\sim_1\{\mt{0}^{n+1},\mt{1}^{n+1}\}$ because
		$(\mt{0}^n,\mt{1}^n)\approx_1 (\mt{0}^{n+1},\mt{1}^{n+1})$.

		One way to prove that $\dim_{\VC}({\NFA'_2(q)}\cap B_n)\le \dim_{\VC}({\NFA'_2(q)}\cap B_{n+1})$ would be if
		each shattered set of words of length $n$ is similar to a shattered set of words of length $n+1$.
		Things turn out not to be that simple:
		\begin{proposition}[{\cite{KT}}]\label{hilo}
			For all $S\subseteq B_3$ with $\abs{S}=3$,
			$\{\mt{00},\mt{01},\mt{10}\}\not\sim_2 S$. 
		\end{proposition}
		\Cref{hilo} says that for any such $S$, the number of subsets $F\subseteq [3]$ for which a suitable automaton $M$ exists is always smaller than 8.
		In fact, our computer calculation shows that the greatest number of sets $F$ for which $M$ exists is 6, which is achieved
		for $S=\{\mt{000},\mt{001},\mt{100}\}$.

		Some weak support for \Cref{cocoaRef} is given by \Cref{thm:support}.

		\begin{theorem}\label{thm:support}
			For all $q\ge 1$ and $n\ge 0$,
			\[
				\dim_{\VC}({\NFA'_{\mathbb N}(q)}\cap B_n) \le \dim_{\VC}({\NFA'_{\mathbb N}(q)}\cap B_{n+1}).
			\]
		\end{theorem}
		\begin{proof}
			Suppose $\dim_{\VC}({\NFA'_{\mathbb N}(q)}\cap B_n) \ge s$ for $s\in\mathbb N$.
			Thus there exists a set $S\subseteq [b]^*$, $\abs{S}=s$ of words of length $n$ that can be shattered by $q$-state NFAs over a finite alphabet $\Sigma\subset\mathbb N$.
			We may assume the NFAs only have transitions for the symbols occurring in words in $S$.
			Let $b\in\mathbb N$ where $b$ does not occur in any word in $S$.
			Let $S'=\{bx:x\in S\}$. We shall show how to shatter $S'$.

			Fix any sets of words $A\subseteq S$, $R\subseteq S$, with $A\cap R=\emptyset$.
			Let $A'=\{bx: x\in A\}$ and $R'=\{bx: x\in R\}$.
			By assumption, there exists $M\in\nfa_{\Sigma}(q)$ with $A\subseteq L(M)$ and $L(M)\cap R=\emptyset$.
			Thus, $M$ accepts all the words in $A$ and rejects all the words in $R$.

			Let $M'$ be formed from $M$ by adding a loop labeled $b$ to the start state $q_0$.
			Then $A'\subseteq L(M')$, $R'\cap L(M')=\emptyset$.
			Thus the automata $M'$ witness that $\dim_{\VC}({\NFA'_{\mathbb N}(q)}\cap B_{n+1})\ge s$, as desired.
		\end{proof}

		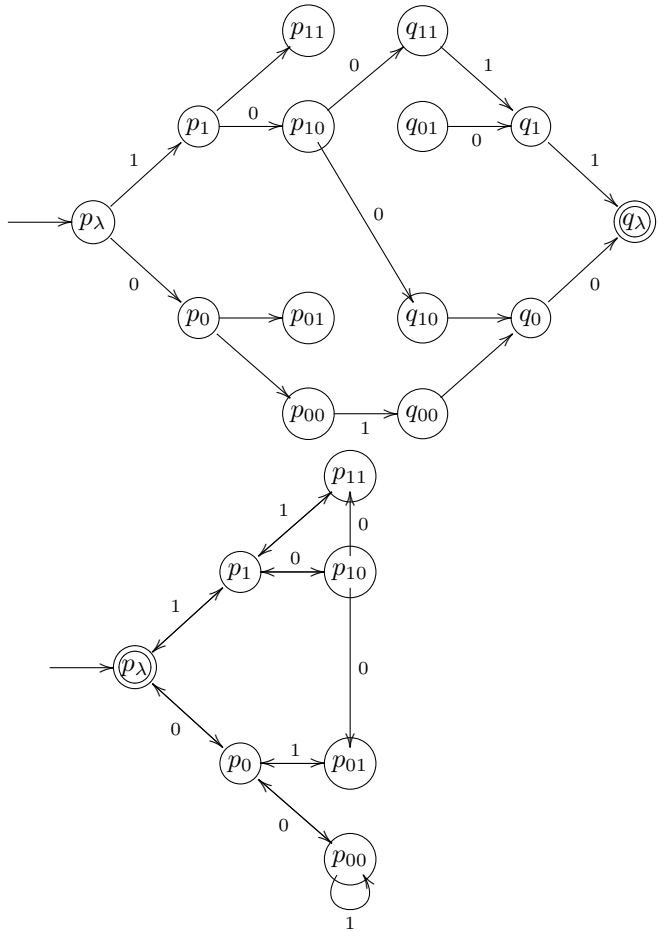
\begin{figure}
			\[
				\xymatrix{
			&												&								& *+[Fo]{p_{11}}					& *+[Fo]{q_{11}}\ar[dr]^1\\
			&												& *+[Fo]{p_1}\ar[r]^{0}\ar[ur] 	& *+[Fo]{p_{10}}\ar[ddr]^0\ar[ur]^0 & *+[Fo]{q_{01}}\ar[r]_0	& *+[Fo]{q_1}\ar[dr]^1\\
			\ar[r]&	*+[Fo]{p_{\lambda}}\ar[dr]_{0}\ar[ur]^{1} 	&								&							 		&							& & *+[Foo]{q_{\lambda}}\\
			&												& *+[Fo]{p_0}\ar[dr]	\ar[r]	& *+[Fo]{p_{01}}					& *+[Fo]{q_{10}}\ar[r]		& *+[Fo]{q_0}\ar[ur]_0\\
			&												&								&*+[Fo]{p_{00}}\ar[r]_1				& *+[Fo]{q_{00}}\ar[ur]
				}
			\]
			\[
				\xymatrix{
			&	&& *+[Fo]{p_{11}}\ar[dl] & \\
			&												& *+[Fo]{p_1}\ar[r]^{0}\ar[ur]^1\ar[dl] 	& *+[Fo]{p_{10}}\ar[u]_0\ar[dd]^0\ar[l] \\
			\ar[r]&	*+[Foo]{p_{\lambda}}\ar[dr]_{0}\ar[ur]^{1} 	&									& 		&  & & \\
			&												& *+[Fo]{p_0}\ar[dr]_0\ar[ul]\ar[r]^1		& *+[Fo]{p_{01}}\ar[l]\\
			&												&&*+[Fo]{p_{00}}\ar[ul]\ar@(dl,dr)_{1}
				}
			\]
			\caption{The Gruber--Holzer construction for length $n=5$ and the language $L=\{10011, 10010,00100\}$, and our modified construction.
			}
		\end{figure}
	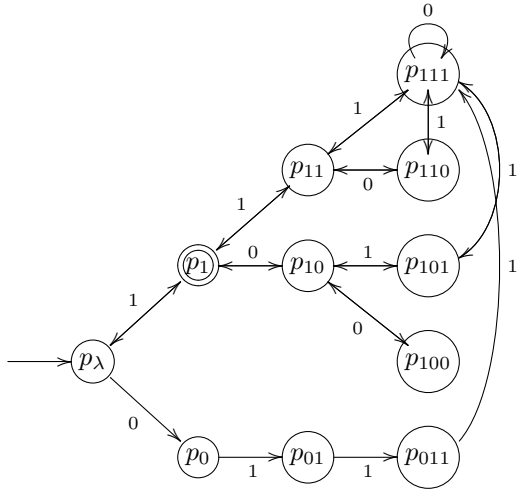
\begin{figure}
			\[
				\xymatrix{
&															&										&										& *+[Fo]{p_{111}}\ar[dl]\ar@(ul,ur)^0\ar[d]\ar@(r,r)[dd]\\
&															&										&*+[Fo]{p_{11}}\ar[dl]\ar[ur]^1\ar[r]_0	& *+[Fo]{p_{110}}\ar[l]\ar[u]_1\\
&															& *+[Foo]{p_1}\ar[r]^{0}\ar[ur]^1\ar[dl] & *+[Fo]{p_{10}}\ar[r]^1\ar[dr]_0\ar[l] & *+[Fo]{p_{101}}\ar[l]\ar@(r,r)[uu]_1\\
\ar[r]&				*+[Fo]{p_{\lambda}}\ar[dr]_{0}\ar[ur]^{1} 	&										& 										& *+[Fo]{p_{100}}\ar[ul]\\
&															& *+[Fo]{p_0}\ar[r]_1					& *+[Fo]{p_{01}}\ar[r]_1 			& *+[Fo]{p_{011}}\ar@(r,r)[uuuu]_1\\
				}
			\]
		\caption{
			The even-length construction in \Cref{thm:HGH} examplified at $n=6$ with $q=11$ states.
			Various layouts of transitions here will shatter $S=\{w\in B_6: 011\preceq w\text{ or }1\preceq w\}$, $\abs{S}=40$ (\Cref{tab:vc-td}).
		}\label{fig:HGH-even}
	\end{figure}
		\begin{figure}
			\[
				\xymatrix{
			&												&										&										& *+[Fo]{p_{111}}\ar@(ul,ur)^0\ar[d]\ar@(r,r)[dd]\ar@(r,r)[dddd]\\
			&												&										&*+[Fo]{p_{11}}\ar[dl]\ar[ur]^1\ar[r]_0	& *+[Fo]{p_{110}}\ar[u]_1\\
			&												& *+[Fo]{p_1}\ar[r]^{0}\ar[ur]^1\ar[dl] & *+[Fo]{p_{10}}\ar[r]^1\ar[dr]_0\ar[l] & *+[Fo]{p_{101}}\ar@(r,r)[uu]_1\\
			\ar[r]&	*+[Foo]{p_{\lambda}}\ar[dr]_{0}\ar[ur]^{1} 	&										& 										& *+[Fo]{p_{100}}\\
			&												& *+[Fo]{p_0}\ar[r]_1					& *+[Fo]{p_{01}}\ar[l]\ar[r]_1 			& *+[Fo]{p_{011}}\ar[l]\ar@(r,r)[uuuu]_1\\
				}
			\]
			\caption{The odd-length construction in \Cref{thm:HGH}. The states shown here suffice to shatter the set $S$ of all words of length 7 that do not begin or end with 000, 001, or 010.
			In particular, the automaton $M$ shown here has $L(M)\cap S=S'=\{01^6,101^5,1101^4,1^301^3,1^4011, 1^501, 1^60\}$.}\label{fig:HGH-odd}
		\end{figure}
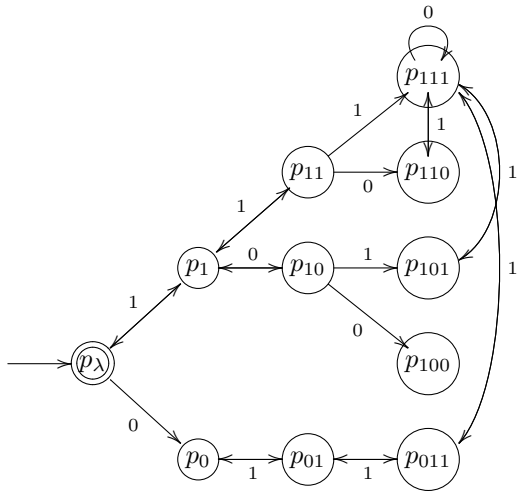
		\begin{table}
			\centering
			\begin{tabular}{c|c|c|c|c|c| c| c| c c c c}
				 $q$ \\
				 \\
				 $\infty$ & 1 & 2 & 4 & 8 & $2^4$		& $2^5$ & $2^6$ & \dots \\
				 \dots	  & \dots & \dots & \dots & \dots & \dots & \dots & \dots & \dots\\
				 15 	  & 1 & 2 & 4 & 8 & $2^4$			& $2^5$		& $\mathbf{2^6}$& $\mathbf{2^7}$ \\ \hline
				 14 	  & 1 & 2 & 4 & 8 & $2^4$			& $2^5$		& \mc{$\{\ge 56\}$\\ /?}	& \mc{$\{\ge 98\}$\\ /?} \\ \hline
				 13		  & 1 & 2 & 4 & 8 & $2^4$			& $2^5$		&					&\\ \hline
				 12		  & 1 & 2 & 4 & 8 & $2^4$			& $2^5$		& \mc{$\{\ge 48\}$\\ /?}	& \mc{$\{\ge 72\}$\\ /?} \\ \hline
				 11		  & 1 & 2 & 4 & 8 & $2^4$			& $2^5$		& \mc{$\{\ge 40\}$\\ /?}	& \mc{$\{\ge 50\}$\\ /?} \\ \hline
				 10		  & 1 & 2 & 4 & 8 & $2^4$			& $2^5$		&   &\\ \hline
				  9		  & 1 & 2 & 4 & 8 & $2^4$ 		& $2^5$		&   &\\ \hline
				  8		  & 1 & 2 & 4 & 8 & $2^4$ 		& $2^5$ 		& \mc{$\{\ge 32\}$\\ /?}  & \mc{$\{\ge 32\}$\\ /?} \\ \hline
				  7		  & 1 & 2 & 4 & 8 & $\mathbf{2^4}$	& $\mathbf{2^5}$	& \mc{$\{\ge 18\}$\\ /?}\\ \hline
				  6		  & 1 & 2 & 4 & 8 & \mc{$\ge 12$\\ /?}	& \mc{$\{\ge 18\}$\\ /?} & ?\\ \hline
				  5		  & 1 & 2 & 4 & 8 & ? 					& 				?	& \mc{$\{\ge 8\}$\\ /?}\\ \hline
				  4		  & 1 & 2 & 4 & 8 & \mc{$\ge 10$\\ /?} & ? & ?\\ \hline
				  3		  & 1 & 2 & $\mathbf{2^2}$ & $\mathbf{2^3}$ & 9*/5* & \mc{$\ge 8$\\ /4} & \mc{$\ge 8$\\ /$[2,4]$} & \mc{($\ge 7$)\\ /$[2,4]$}\\ \hline
				  2		  & 1 & 2 & 4/4 & 5/3 & 5/2				& 5/2 & $\ge 5$/1 & $\ge 5$/1\\
				 \hline
				  1		  & $\mathbf{2^0}$ & $\mathbf{2^1}$ & 2/1 & 2/1 & 2/1 & 2/1 & 2/1 &2/1 & 2/1\\
				 \hline
				 & $0$ & $1$ & $2$ & $3$ & $4$ & $5$ & $6$ & 7&$\infty$ & $n$ \\
			\end{tabular}
		\caption{
		Cells annotated with $a/b$ or just $a$ indicate $a=\dim_{\VC}({\NFA'_2(q)}\cap B_n)$ or $b=\dim_{\TD}(\NFA'_2(q)\cap B_n)$.
		When these coincide the value $a$ may be indicated.
		For $n\ge 3$ and $q\ge 2$ see \cite{KT}.
		Numbers in \{braces\} come from \Cref{thm:HGH}.
		\textbf{Bold} numbers indicate results derived from \Cref{thm:HGH} with $r=0$.
		* = result also correct for $\NFA$ in place of $\NFA'$ (all numbers for $n\le 3$ or $q\le 2$ are also correct for $\NFA$ in place of $\NFA'$).
		}\label{tab:vc-td}
		\end{table}

\bibliographystyle{plain}
\bibliography{vc}
\end{document}